\documentclass[conference]{IEEEtran}

\IEEEoverridecommandlockouts                              

\usepackage{microtype}
\usepackage{graphicx}
\usepackage{subcaption}
\usepackage{booktabs} 
\usepackage[ruled,linesnumbered]{algorithm2e}
\usepackage{float}
\usepackage{graphicx}
\usepackage{textcomp}

\usepackage{amsfonts,amsmath,amssymb,amsthm,url,algorithmic}
\newtheorem{remark}{Remark}
\newtheorem{theorem}{Theorem}

\def\BibTeX{{\rm B\kern-.05em{\sc i\kern-.025em b}\kern-.08em
    T\kern-.1667em\lower.7ex\hbox{E}\kern-.125emX}}

\usepackage{balance}
    
\begin{document}

\title{Distributed and Efficient Resource Balancing\\ Among
  Many Suppliers and Consumers}
  
\author{\IEEEauthorblockN{Kamal Chaturvedi}
\IEEEauthorblockA{
kamal.chaturvedi@colorado.edu}
\and
\IEEEauthorblockN{Jia Yuan Yu}
\IEEEauthorblockA{
jiayuan.yu@concordia.ca}
\and
\IEEEauthorblockN{Shrisha Rao}
\IEEEauthorblockA{
shrao@ieee.org}
}
\maketitle

\begin{abstract}

Achieving a balance of supply and demand in a multi-agent system with
many individual self-interested and rational agents that act as
suppliers and consumers is a natural problem in a variety of real-life
domains---smart power grids, data centers, and others.  In this paper,
we address the profit-maximization problem for a group of distributed
supplier and consumer agents, with no inter-agent communication. We
simulate a scenario of a market with $S$ suppliers and $C$ consumers
such that at every instant, each supplier agent supplies a certain
quantity and simultaneously, each consumer agent consumes a certain
quantity.  The information about the total amount supplied and
consumed is only kept with the center.  The proposed algorithm is a
combination of the classical additive-increase multiplicative-decrease
(AIMD) algorithm in conjunction with a probabilistic rule for the
agents to respond to a capacity signal.  This leads to a
nonhomogeneous Markov chain and we show almost sure convergence of
this chain to the social optimum, for our market of distributed
supplier and consumer agents. Employing this AIMD-type algorithm, the
center sends a feedback message to the agents in the supplier side if
there is a scenario of excess supply, or to the consumer agents if
there is excess consumption. Each agent has a concave utility function
whose derivative tends to 0 when an optimum quantity is
supplied/consumed.  Hence when social convergence is reached, each
agent supplies or consumes a quantity which leads to its individual
maximum profit, without the need of any communication.  So eventually,
each agent supplies or consumes a quantity which leads to its
individual maximum profit, without communicating with any other
agents.  Our simulations show the efficacy of this approach.

\end{abstract}


\section{Introduction}

Many modern problems involve one resource that is generated over time
by a number of suppliers and simultaneously consumed by a number of
consumers.  For instance, consider electrical power generated or
injected into the grid by utility companies, wind mills, solar panels,
batteries, and electric vehicles. At the same time, other devices
consume the same power.  Other examples include cryptocurrency markets
with large numbers of producers and consumers, etc.

These problems are complex problems that involve a huge number of
decision-makers. Each producer has a different cost for producing a
unit of the resource, and each consumer derives a different benefit
from consuming the resource.  One desirable outcome of the interaction
between producers and consumers is to maximize the social welfare or
minimize the social cost.  If we can represent each agent by an
utility function, then we aim to maximize the sum of utilities of
producers and consumers subject to capacity constraints on the
quantities. These constraints ensure that quantities are nonnegative
and that the total consumption is at most the total production.

It is well-known that in the presence of money and price signals, an
equilibrium can be reached through a tatonnement process where social
welfare is maximized.  However, we propose a mechanism where neither
money, nor price signals, are required to maximize social welfare: we
only require binary feedback signals on whether or not total
consumption exceeds total production. Our mechanism is decentralized:
each agent only requires knowledge of its own utility function. We
show that if every agent follows a specified algorithm, then over
time, the profile of decisions (production and consumption quantities)
converges to a socially optimal outcome.

This problem can be related to the TCP congestion control problem,
where data sources controlling their own rates can interact to achieve
an optimal network-wide rate allocation, where supplier agents can act
as source of allocation, while consumer agents act as the consumers of
the allocated resources.  The detailed
analysis~\cite{chiu1989analysis} of increase/decrease algorithms used
predominantly in the past for TCP congestion control, which include
Multiplicative Increase/Multiplicative Decrease
(MIMD)~\cite{altman2005analysis}, Additive Increase/Additive Decrease
(AIAD)~\cite{xu2005stability}, Additive Increase/ Multiplicative
Decrease (AIMD)~\cite{yang2000general}, and Multiplicative
Increase/Additive Decrease (MIAD), shows that AIMD leads to the most
optimal resource allocation.

AIMD as a feedback control algorithm has been investigated a lot in
the
literature~\cite{jacobs1996providing,rejaie1999rap,sisalem1998loss,cen1997flow,corless2016aimd}.
In this setting, each agent determines its own allocation as per the
AIMD algorithm.  Similar distributed optimization algorithms have been
shown to iteratively converge to an optimal allocation of
resources~\cite{ram2010distributed,nedic2009distributed,duchi2012dual},
but these algorithms rely on inter-agent communication to achieve
optimality.  While a combined version of the classical AIMD algorithm
with a probabilistic rule \cite{aimd} defines the
behavior of multiple agents in response to a capacity signal. This
non-homogeneous Markov chain is showed to reach sure convergence, to
the social optimum. Here, each agent responds to a capacity event
according to its own probability function, known only to that agent,
without the need of any communication with another agent. In this
sense,~\cite{aimd} goes beyond traditional AIMD and
emulates RED-like congestion control \cite{srikant2004internet}. Also,
very limited actuation is assumed for an agent, and the agent only
decides to respond to a capacity event or not, in an asynchronous
manner. There is no need for a common clock and the setting is
completely stochastic.


More generally, problems concerning multiple independent producers and
consumers can occur in many different domains of application---for
instance, electrical power~\cite{muralidharan2018}, solar energy and
microgrids~\cite{maity2010}, data centers~\cite{nsingh2014}, and air
cargo~\cite{totamane2012}.  Such problems can be considered as
specific instances of machine learning~\cite{mitchell1997}, or as
problems of collective control over distributed dynamical
systems~\cite{wolpert2000}.

A simple strategy of \emph{rational learning} where each producer and
each consumer acts based on the previous instance, is
unsatisfactory~\cite{enumula2009}.  Many classes of equilibrium
problems are computationally hard~\cite{codenotti2008}.


This paper is organized as follows. First, we describe the
mathematical model in Section~\ref{sec-setting}.  We describe our
proposed solution approach in Section~\ref{sec-algorithm}.  We show
the optimality of our approach in Section~\ref{sec-conv} and describe
the iterative convergence to optimality through simulations in
Section~\ref{sec-sim}.  Finally, we sum up our contributions and
conclude in Section~\ref{sec-conclude}.

\section{Setting}\label{sec-setting}

We consider a market with a single good (arbitrarily divisible), with
multiple suppliers and multiple consumers. Three types of convergence
are routinely observed.  First, the total demand and total supply
converge through a dynamic process that requires the exchange of
information on the price of a good, the total demand, or the total
supply (e.g., through a process of tatonnement).  In the presence of
multiple suppliers, these suppliers may compete via their production
quantities, which is known as Cournot competition. In Cournot
competition, the production quantity of each supplier may converge as
well through a tatonnement process.  Similarly, in the presence of
multiple consumers, these consumers may also compete for the limited
quantity of good available through bidding and proportional allocation
(e.g., competition among newsvendors). Here again, we observe
convergence of the allocated quantities of each consumer.  In these
situations, the convergence relies on an exchange of information in
the form of prices, bids, and quantities of goods.  In this work, we
consider a situation where communication, e.g., the exchange of
information, is limited.

The notation that we use throughout the paper is as follows. There is
a group of suppliers $S$, and a group of consumers $C$.  Time is
discrete: $t=1,2,\ldots$ For each $i\in S$, the amount of good
supplied at time $t$ is $x_i(t)$.  For each $j\in C$, the amount of
good consumed at time $t$ is $y_j(t)$.  The total supply of good at
time $t$ is $x(t)$, the total consumption is $y(t)$.  For each $i\in
S$, the utility function $f_i(z)$ represents the profit from producing
an amount $z$.  For each $j\in C$, the utility function $g_j(z)$
represents the profit from consuming an amount $z$.  Our objective is
twofold. On one hand, we want to see the individual consumption
quantities and production quantities to converge using limited
inter-agent communication. On the other hand, we want the limit to be
efficient: the total supply and total demand are balanced, the total
supply is allocated optimally among suppliers, and the total demand is
allocated optimally among the consumers.

\begin{remark}[Choice of utility function]
Whether the agent wants to respond to the capacity signal or not,
highly depends on the choice of the utility function for that
agent. The utility function has to be selected, considering the fact
that it should lead to a maximum so that we can reach an optimal
point, which maximizes profit for the agent. Therefore, we consider
concave functions, which will be demonstrated to reach a global
maximum at the optimum quantity supplied/consumed, in the later
sections.
\end{remark}

More precisely, the first objective requires that there exist
constants $\{x_i^*\}$ and $\{y_i^*\}$ such that for every $i$ and $j$,
we have
\begin{align}
    x_i(t) \to x_i^*,\\
    y_j(t) \to y_j^*,
\end{align}
as $t\to \infty$. The second objective requires that $\{x_i^*\}$
and $\{y_i^*\}$ satisfy the following:
\begin{align}
    (x^*, y^*) \in \max_{x,y} \quad &\sum_{i\in S} f_i(x_i) +
                                      \sum_{j\in C} g_j(y_j)\\
    \mbox{such that} \quad &\sum_{i\in S} x_i = \sum_{j\in C} y_j.
\end{align}
The limited communication property
will be presented after presenting the algorithms for computing
$\{x_i(t),y_j(t)\}$.  

For the sake of discussion, and as a baseline, we also define the
following vectors $u^*$ and $w^*$ solving the following
\emph{unconstrained} optimization problems:
\begin{align}\label{u}
   u^* \in \arg\max_{x \in \mathbb R^S} \quad &\sum_i f_i(x_i),
\end{align}
and
\begin{align}\label{w}
   w^* \in \arg\max_{y \in \mathbb R^C} \quad &\sum_j g_j(y_j).
\end{align}


Next, we present the proposed distributed algorithm that specifies how
agents update their $x_i(t)$ and $y_j(t)$ over time.

\section{Algorithm}\label{sec-algorithm}

We consider a distributed environment, where we have a total of S
suppliers and C consumers, such that each time instant $t$, each
supplier $i$ supplies an amount $x_i(t)$ and after that, each consumer
$j$ consumes an amount $y_j(t)$.

The procedures for updating $x_i(t)$ and $y_j(t)$ are presented in
Algorithm~\ref{algos} and Algorithm~\ref{algoc}.

At the time instant $t$, we keep a check whether the total amount
supplied at $(t-1)$ was equivalent to the total amount consumed, or
not. If the supply was more than the consumption, a capacity signal is
sent to all the supplier agents to reduce the supply amount for
$t$. Otherwise, if the consumption was more than the supply, a
capacity signal is sent to the consumer agents to reduce the
consumption for $t$. This makes sure that an equilibrium is maintained
throughout the execution, whenever the supply or consumption
fluctuates.

If the agent responds to the capacity signal, it reduces the amount in
the next time instant by a factor of $\beta$, or else, it keeps on
increasing the amount after every time instant by adding the value of
$\alpha$ to it. When the agent receives a capacity signal, its probability to reduce by a factor of $\beta$ depends on the Bernoulli random variable $b_i(t)$, as
$P(b_i(t) = 1) = \lambda$, and  $P(b_i(t) = 0) = 1 -\lambda$

An agent also keeps track of its individual long term
average, which is given by:
\begin{equation*}
\bar{x_i}(t)=\Big(\frac{1}{t+1}\Big)\sum_{T=0}^{t} x_i(T).
\end{equation*}

We want to consider limited inter-agent communication. The only
communication available at time $t$ are the signals:
\begin{align}
    s(t) & \triangleq 1_{[x(t-1) < y(t-1)]} \nonumber\\
  &= 1_{[\sum_i x_i(t-1) < \sum_j y_j(t-1)]} \quad t=1,2,\ldots
\end{align}

\begin{algorithm}
\caption{AIMD Algorithm Supplier $i$}
\label{algos}
\begin{algorithmic}[1]
\STATE {\bfseries Input:} $t$, $f_i$,$x_i(t-1)$, $s(t)$, $\Gamma$, $\alpha_s$ and $\beta_s$
\STATE {\bfseries Output:} $x_i(t)$
\STATE {\bfseries Initialization:} $\lambda=0$ and $\bar{x_i}(t)=0$
\IF{($s(t)=1$)\label{algo:line:one}}
\STATE{$\lambda = \Gamma (f^{\prime}_i(\bar{x_i}(t-1)) /\bar{x_i}(t-1))$}\label{algo:line:two}\
\STATE{generate independent Bernoulli random variable $b_i(t)$ with parameter $\lambda$}\label{algo:line:three}\
\ENDIF
\IF{($b_i(t)=1$)}
\STATE{$x_i(t)=x_i(t-1) \beta_s$}\label{algo:line:four}\
\ELSIF{$x_i(t-1) \leq {x_i}^*$ \label{algo:line:five}}
\STATE{$x_i(t)=x_i(t-1)+\alpha_s$}\label{algo:line:six}\
\ELSIF{$x_i(t-1)>{x_i}^*$ \label{algo:line:seven}}
\STATE{$x_i(t) =x_i(t-1)-\alpha_s$}\label{algo:line:eight}\
\ENDIF
\STATE{$\bar{x_i}(t) = ((\bar{x_i}(t-1) \cdot (t- 1))+x_i(t)) / (t+1)$}\label{algo:line:nine}\;
\RETURN{$x_i(t)$\label{algo:line:ten}}
\end{algorithmic}
\end{algorithm}

\begin{algorithm}
\caption{AIMD Algorithm Consumer $j$}
\label{algoc}
\begin{algorithmic}[1]
\STATE {\bfseries Input:} $t$, $g_j$,$y_j(t-1)$, $c(t)$, $\Gamma$, $\alpha_c$ and $\beta_c$
\STATE {\bfseries Output:} $y_j(t)$
\STATE {\bfseries Initialization:} $\lambda=0$ and $\bar{y_j}(t)=0$
\IF{($c(t)=1$)}\label{algo1:line:one}
\STATE{$\lambda = \Gamma (g^{\prime}_j(\bar{y_j}(t-1)) /\bar{y_j}(t-1))$}\label{algo1:line:two}
\STATE{generate independent Bernoulli random variable $b_j(t)$ with parameter $\lambda$}\label{algo1:line:three}
\ENDIF
\IF{($b_j(t)=1$)}
\STATE{$y_j(t)=y_j(t-1) \beta_c$}\label{algo1:line:four}
\ELSIF{$y_j(t-1) \leq {y_j}^*$}\label{algo1:line:five}
\STATE{$y_j(t) =y_j(t-1)+\alpha_c$}\label{algo1:line:six}
\ELSIF{$y_j(t-1)>{y_j}^*$}\label{algo1:line:seven}
\STATE{$y_j(t) =y_j(t-1)-\alpha_c$}\label{algo1:line:eight}
\ENDIF
\STATE{$\bar{y_j}(t) = ((\bar{y_j}(t-1) \cdot (t- 1))+y_j(t)) / (t+1)$}\label{algo1:line:nine}
\RETURN{$y_j(t)$}\label{algo1:line:ten}
\end{algorithmic}
\end{algorithm}

The pseudocode for the procedure describing what happens at the
supplier side is shown in Algorithm \ref{algos}. Each agent maintains
the last value sent, $x_i(t-1)$ and the long term average of the
values sent in the past, $\bar{x_i}(t)$, and this along with the
concave nature of its utility function helps the agent drive the
quantity sent towards the optimal point, ${x_i}^*$.

At the iteration $t$ of the algorithm, it is first checked whether the
total quantity sent at $(t-1)$ was more than the total quantity
consumed(condition for capacity signal being sent to the agent), or
not (line \ref{algo:line:one}) for each of the supplier agent. If the
condition is satisfied, probability $\lambda$ is calculated (line
\ref{algo:line:two}) as a function of utility derivative and long-term
average, for $(t-1)$. On the basis of this probability, it is
determined whether the agent responds to the capacity signal or
not. The agent responds to the signal by reducing the quantity sent at
$t$ by a factor of $\beta$, in comparison to what it sent at $(t-1)$
(line \ref{algo:line:four}). This depends on the independent Bernoulli
random variable $b_i(t)$ with parameter $\lambda$(line
\ref{algo:line:three}).

Else, it checks whether the value supplied at $(t-1)$ was lesser than
the optimum point of the agent's utility function or not (line
\ref{algo:line:five}). If satisfied, the agent adds the value of
$\alpha$ to quantity supplied at $(t-1)$ and send it at $t$ (line
\ref{algo:line:six}).  Otherwise, it sends the quantity by
subtracting $\alpha$ to the quantity sent at $(t-1)$(line
\ref{algo:line:eight}). This condition is included because we do not
want the agent to over-supply, and also makes sure that if this agent
has reached its optimum supply quantity, other agents follow suit and
are driven towards their individual optimum supply quantities. We will
prove this statement later.

Then, the long term average is calculated (line
\ref{algo:line:nine}), and the quantity is supplied by agent $i$ at
$t$ (line \ref{algo:line:ten}). The same algorithm runs at the
consumer side (Algorithm \ref{algoc}), and both of these concurrent
running algorithms make sure that our system reaches the state which
leads to maximum combined profit.

\section{Convergence}\label{sec-conv}

In this section, we show the convergence over time of the sequences of
produced quantities $\{x_i(t)\}$ for all producers, as well as the
sequences of consumed quantities $\{y_i(t)\}$ for all consumers.

\begin{theorem}[Convergence Theorem]
  Suppose that every function $f_i$ and $g_j$ is concave and achieves
  its maximum at a finite point.  For every producer $i$ and every
  consumer $j$, we have
  \begin{align*}
    x_i(t) \to u_i^*, \quad y_j(t) \to w_j^*.
  \end{align*}
\end{theorem}

\begin{proof}[Proof Sketch]
  Recall $u^*$ and $w^*$ are defined in (\ref{u},\ref{w}).
  The proof proceeds by three steps.
  First, we show that $\sum_j y_j(t)$ converges as $t\to\infty$.
  Secondly, we show that $x_i(t)$ converges for all $i$.
  Lastly, we show that $y_j(t)$ converges for all $j$.

\textbf{Step 0.}  Observe that since each $g_j$ achieves its maximum at
  the point $w_j^*$, hence, there exists a finite number $C_y$ such
  that $\sum_j w_j^* = C_y$. Observe that, if $\sum_j y_j(t)$
  converges, the limit must be $C_y$.

\textbf{Step 1.} Consider the sequence
  \begin{align*}
    \hat s(t) = 1_{[\sum_i x_i(t-1) < C_y]}.
  \end{align*}
  By continuity, the AIMD algorithm with input $\hat s(t)$ and the
  AIMD algorithm with input $s(t)$ converge to the same limit point.
  Therefore, by \cite[Theorem~1]{aimd}, we have
  $x_i(t) \to u_i^*$ for all $i$.

\textbf{Step 2.} Consider the sequence
  \begin{align*}
    \hat c(t) = 1_{[\sum_i u_i^* < \sum_j y_j(t-1)]}.
  \end{align*}
Since $x_i(t) \to u_i^*$ for all $i$, by continuity, the AIMD
algorithm with input $\hat c(t)$ and the AIMD algorithm with input
$c(t)$ converge to the same limit point.  Therefore, by
\cite[Theorem~1]{aimd}, we have $y_j(t) \to w_j^*$
for all $i$.
\end{proof}

\section{Simulations}\label{sec-sim}



\begin{figure}[htbp]
\centering
\begin{subfigure}{.50\textwidth}
\centering
\includegraphics[scale=0.38]{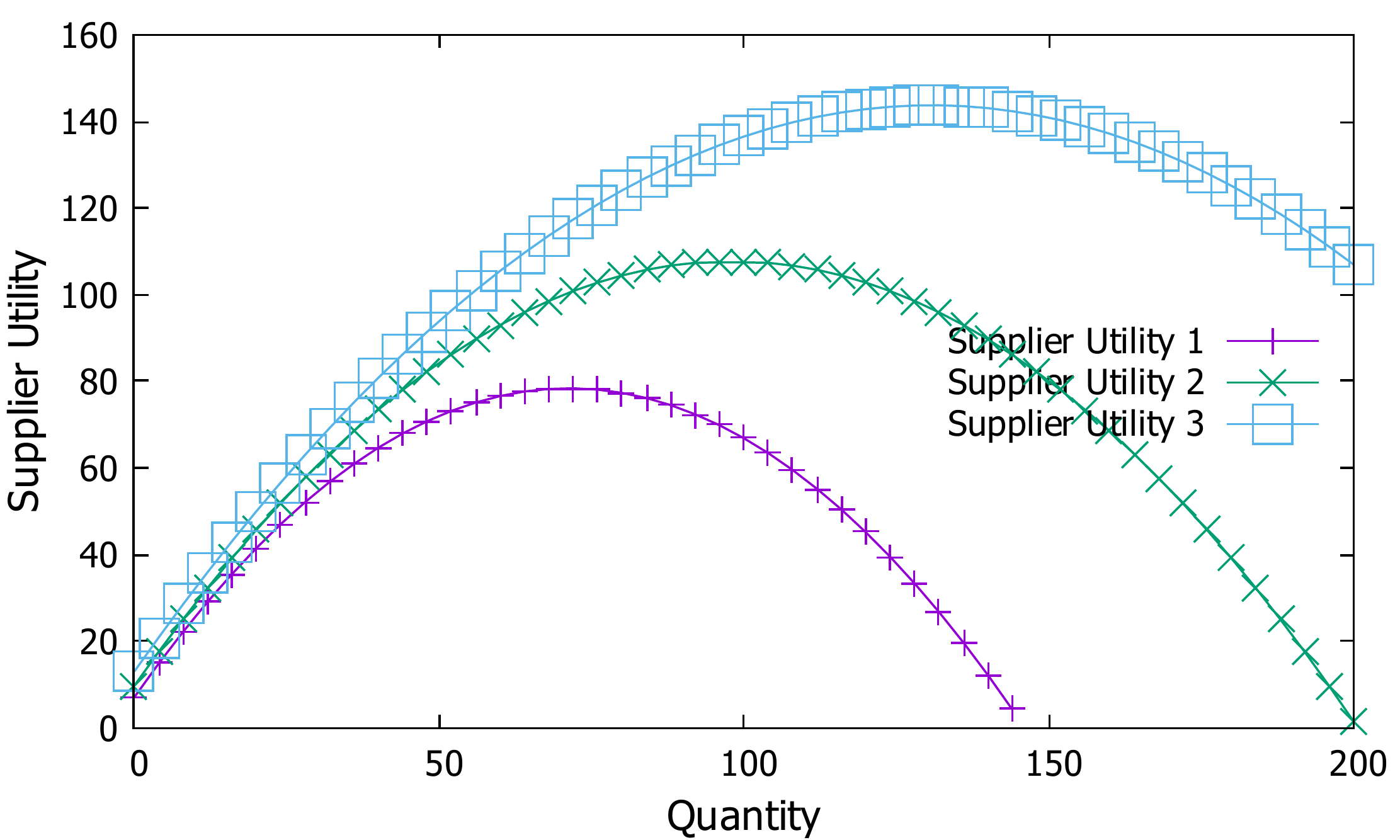}
\caption{Supplier Utility function plot}
\label{fig:9SuppliersUtility}
\end{subfigure}
\begin{subfigure}{.50\textwidth}
\centering
\includegraphics[scale=0.38]{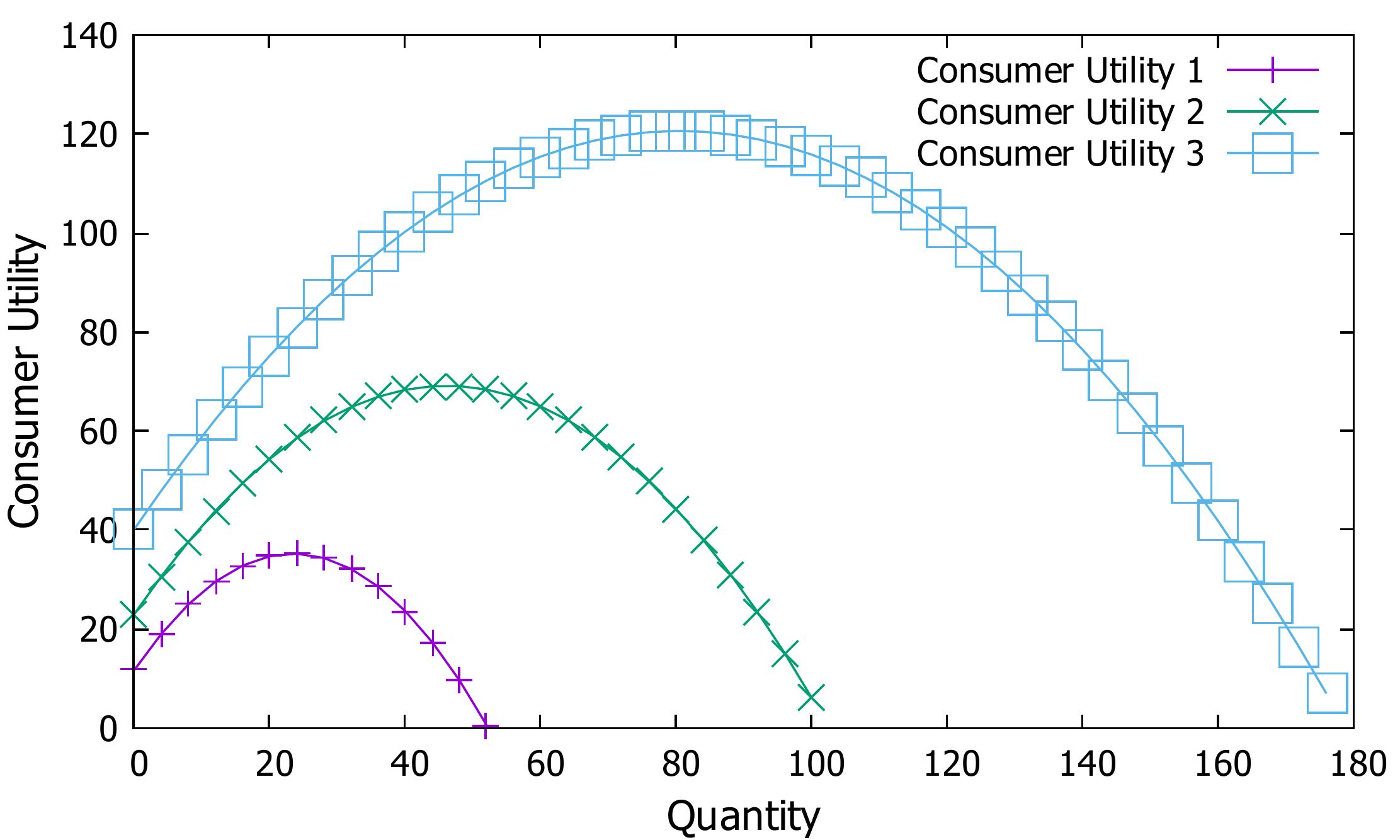}
\caption{Consumer Utility function plot}
\label{fig:18ConsumersUtility}
\end{subfigure}
\label{fig:utilityPlot}
\caption{The concave nature of utility functions}
\end{figure}

In this section, we simulate the interactions of suppliers and consumers in two settings; first, when their utility functions are non-monotonic concave, as in Figure~\ref{fig:9SuppliersUtility}, second, when the supplier utility functions are monotonic concave, as in Figure~\ref{fig:supplierutilitymonotonic}.

\subsection{Non-monotonic Utility Functions}
\label{simulation1}
  We simulate a total of 9 supplier agents, and 18 consumer
    agents. We generate random utility functions $\{f_i, g_j\}$ for the agents, while ensuring that the following sums on the supplier and consumer sides are
    deterministic and equal:
  \begin{equation}
    \sum_{i\in S} \max_{z_i} f_i(z_i) = \sum_{j\in C} \max_{z_j} g_j(z_j) = 900.
    \label{eq:sumOfQuantities}
  \end{equation}
  The values of $\alpha$ and
  $\beta$ for supplier and consumer agents is 5 and 0.75 respectively.
  The network constant $\Gamma$ is kept at 2.0 to ensure that the probability
  $\lambda$ remains in the interval $[0,1]$.

\begin{figure}[htbp]
\centering
\begin{subfigure}{0.45\textwidth}
\includegraphics[scale=0.38]{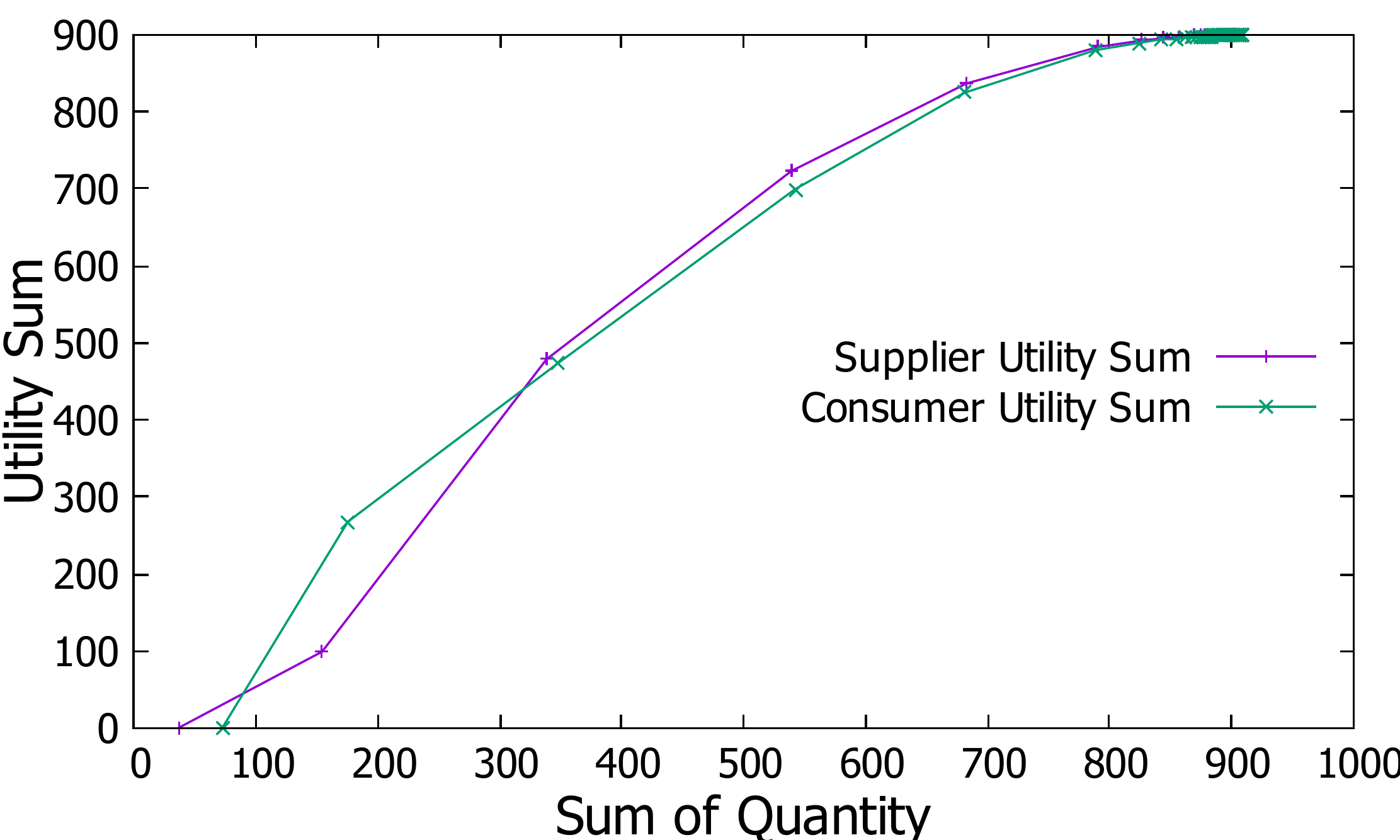}
\caption{Sum of Supplier and Consumer utilities with respect to the change in total supply/consumption}
\label{fig:sumOfUtilities}
\end{subfigure}
\begin{subfigure}{0.45\textwidth}
\includegraphics[scale=0.38]{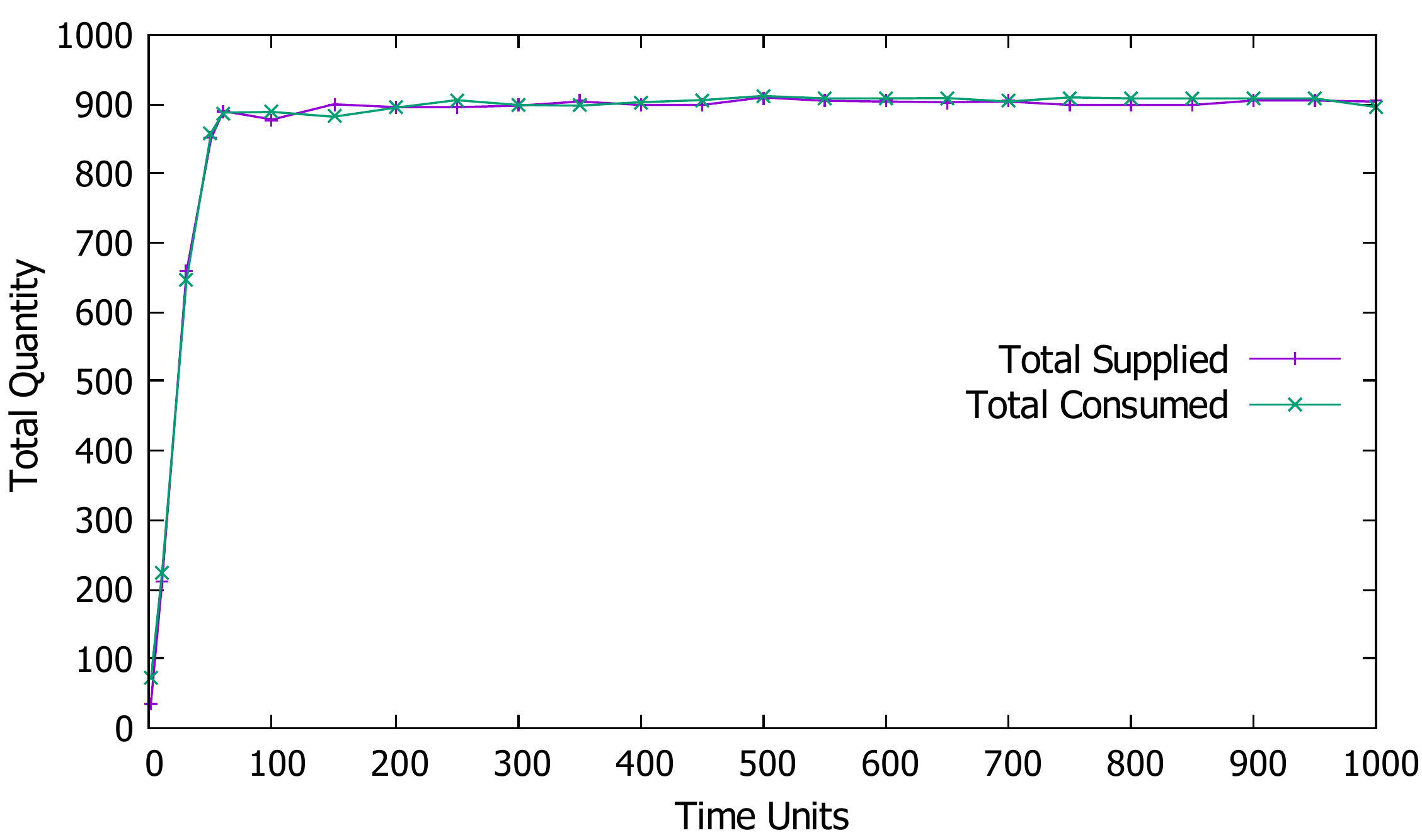}
\caption{Total supply and demand variation with time}
\label{fig:totalsupplyconsumption}
\end{subfigure}
\begin{subfigure}{0.45\textwidth}
\centering
\includegraphics[scale=0.38]{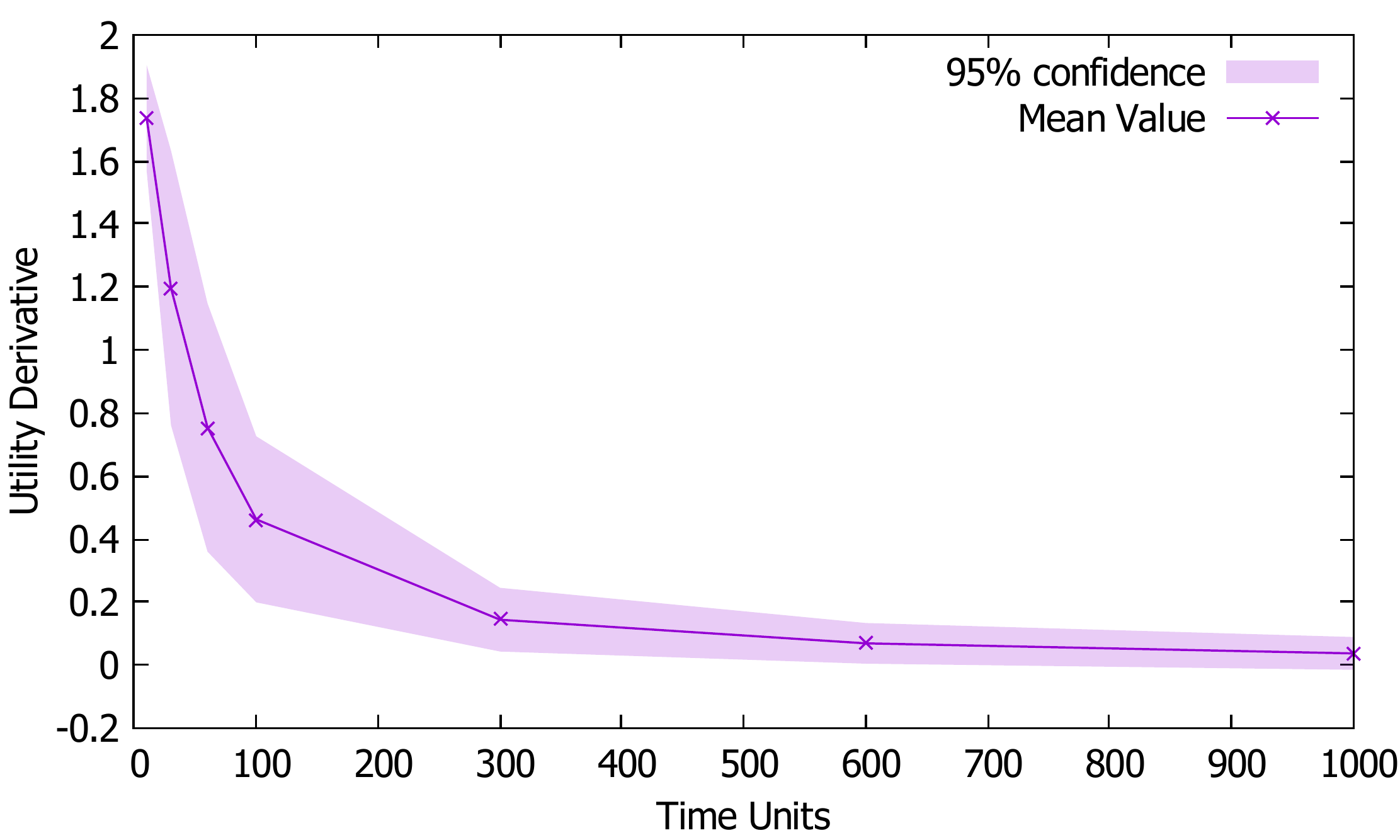}
\caption{Utility Derivative with confidence}
\label{fig:supplierUtilityDerivativeFilled}
\end{subfigure}
\caption{Simulation Analysis for concave supplier and consumer functions}
\end{figure}

\begin{figure}[htbp]
\centering
\begin{subfigure}{.45\textwidth}
\centering
\includegraphics[scale=0.38]{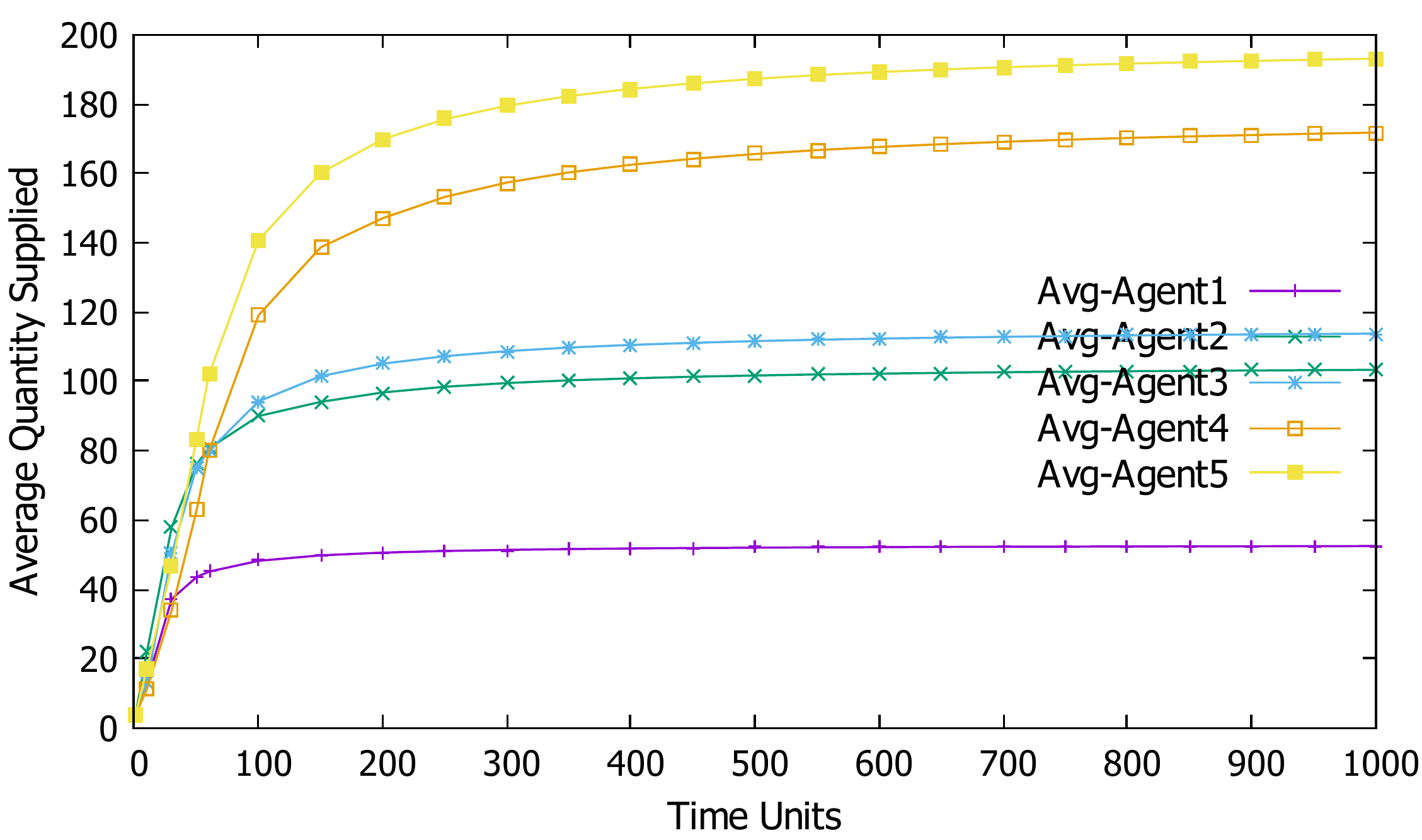}
\caption{Average supplied amount by the 5 agents driven towards their optimum point}
\label{fig:9SuppliersAverage}
\end{subfigure}
\begin{subfigure}{.45\textwidth}
\centering
\includegraphics[scale=0.38]{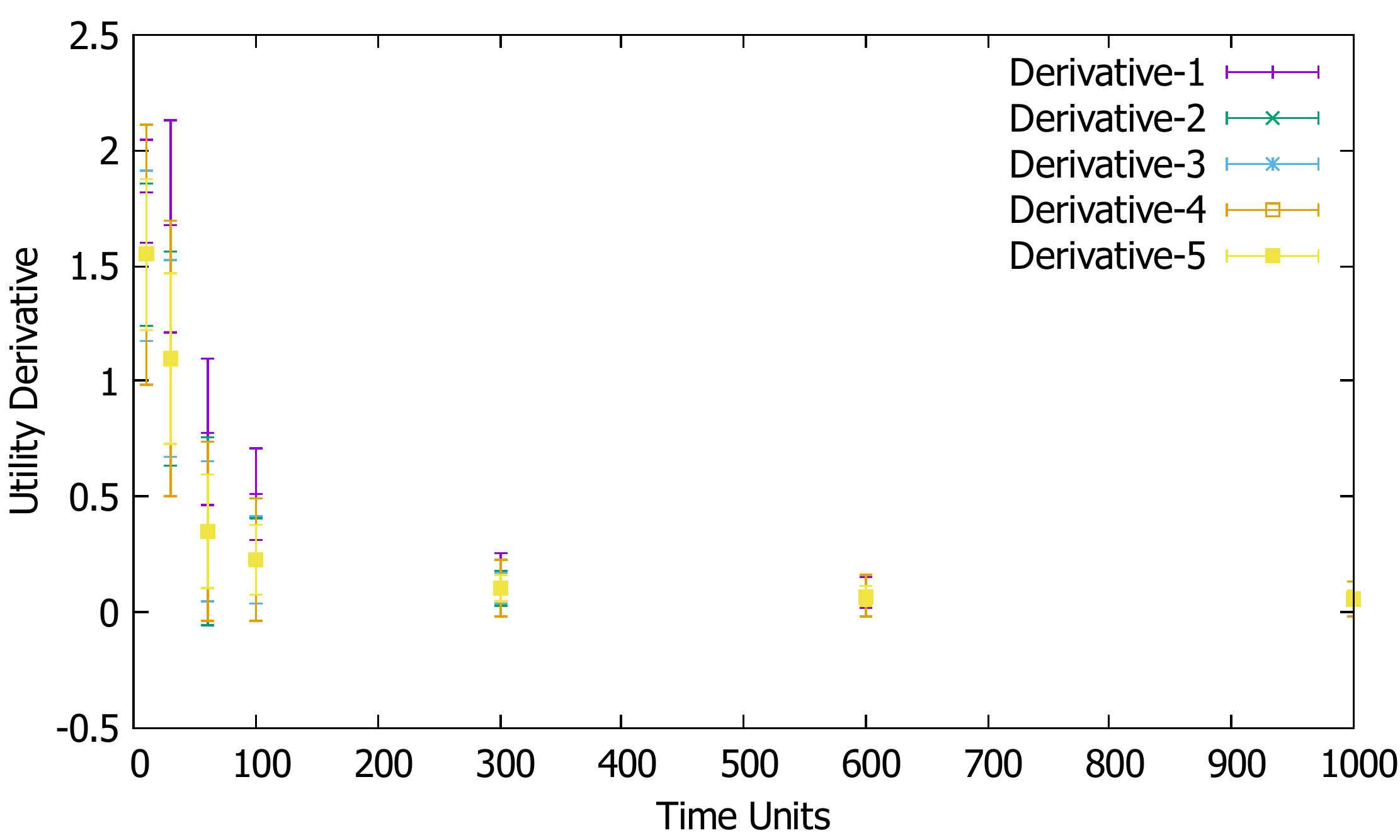}
\caption{Derivative of the utility function of supplier agents converging}
\label{fig:9SuppliersDerivativeFunction}
\end{subfigure}
\label{fig:supplierSideAnalysis}
\caption{Supplier side analysis of algorithm}
\end{figure}

\begin{figure}[htbp]
\centering
\begin{subfigure}{.50\textwidth}
\centering
\includegraphics[scale=0.38]{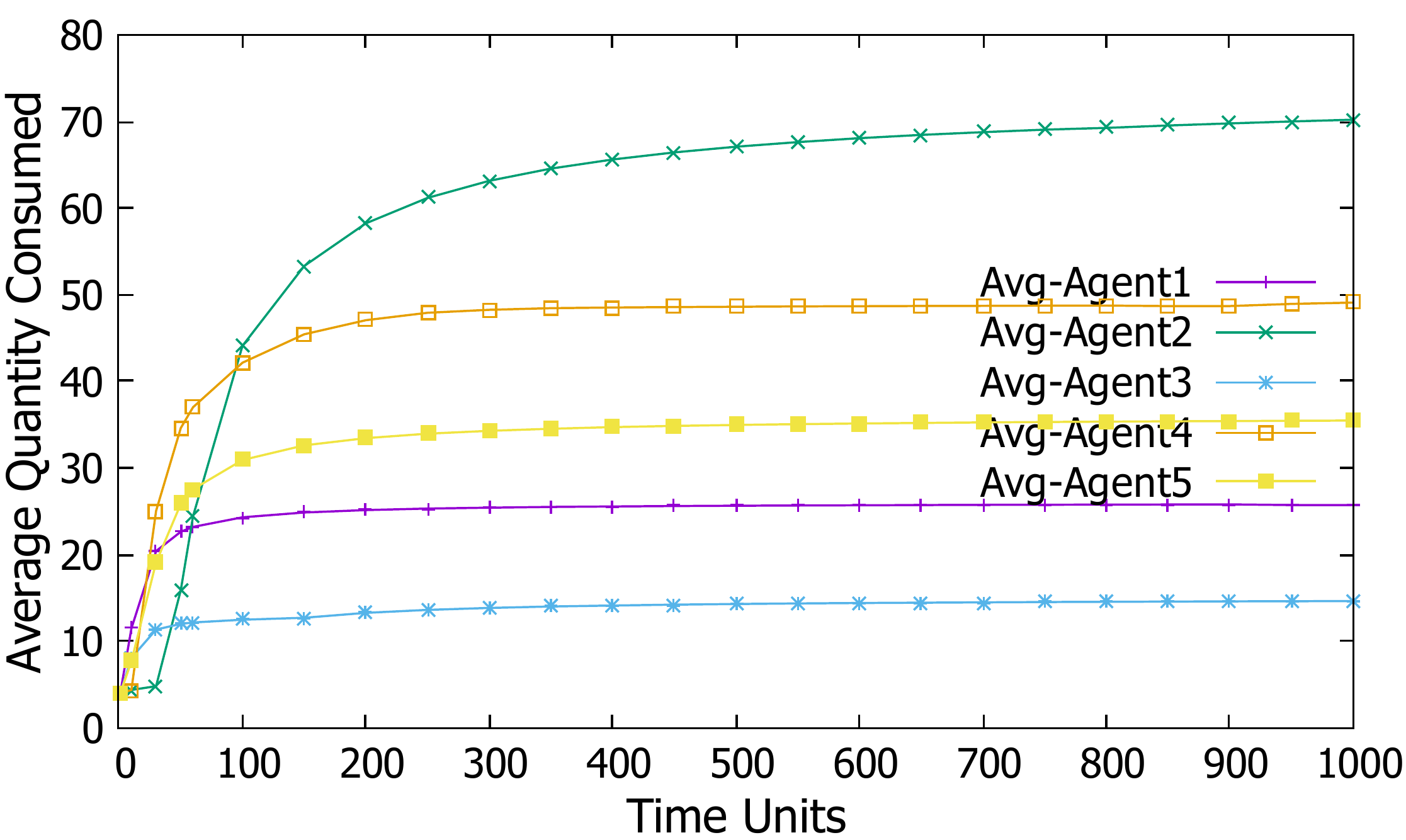}
\caption{Average consumed amount by randomly selected 5 consumer agents, driven towards their optimum point}
\label{fig:18ConsumersAverage}
\end{subfigure}
\begin{subfigure}{.50\textwidth}
\centering
\includegraphics[scale=0.38]{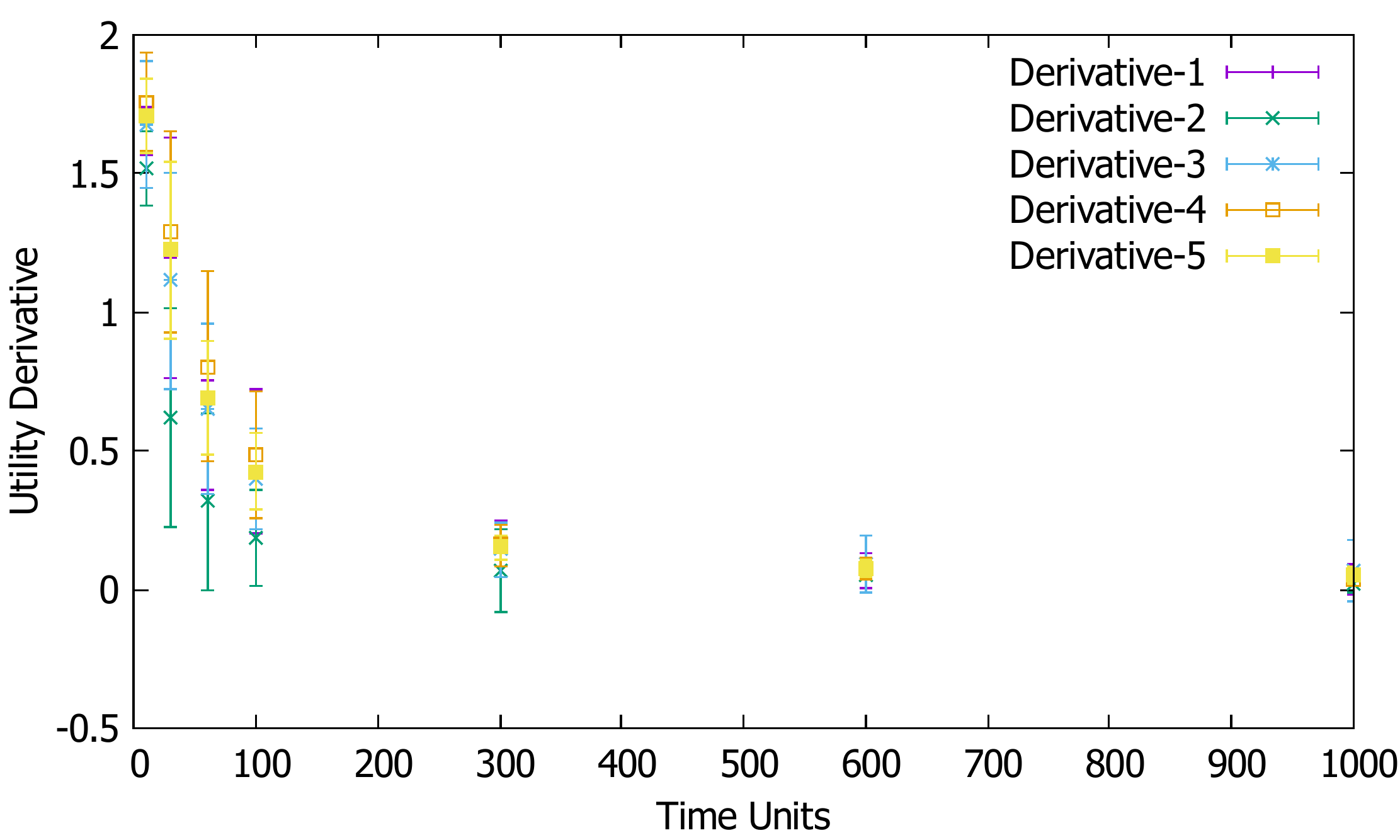}
\caption{Derivative of the utility function of randomly selected 5 consumer agents converging}
\label{fig:18ConsumersDerivativeFunction}
\end{subfigure}
\label{consumerSideAnalysis}
\caption{Consumer side analysis of algorithm}
\end{figure}

The total supply and consumption lingers around the peak total utility, as shown
in (\ref{eq:sumOfQuantities}), i.e. $900$. As seen in
Figure~\ref{fig:totalsupplyconsumption}, the total supply from all
supplier agents is balanced by the total consumption by all consumer
agents.

From Figures~\ref{fig:9SuppliersAverage}
and~\ref{fig:18ConsumersAverage}, we can see that the respective
supplier and consumer agent long-term averages saturate around their
respective optimum points.
Figures~\ref{fig:9SuppliersDerivativeFunction}
and~\ref{fig:18ConsumersDerivativeFunction} show the utility-function
derivatives converging towards 0 for both supplier and consumer
agents, depicting the maximum profit for each agent, while
Figure~\ref{fig:supplierUtilityDerivativeFilled} shows the 95\%
confidence on the supplier utility derivative. The same can be
understood by examining Figure~\ref{fig:sumOfUtilities}, that the sum
of all the utilities converges towards optima, i.e. 900, which was
assumed as a constant in the beginning of our simulation.

\subsection{Monotonic Supplier Utility Functions}
\label{simulation2}
\begin{figure}
\centering
\begin{subfigure}{.50\textwidth}
\centering
\includegraphics[scale=0.38]{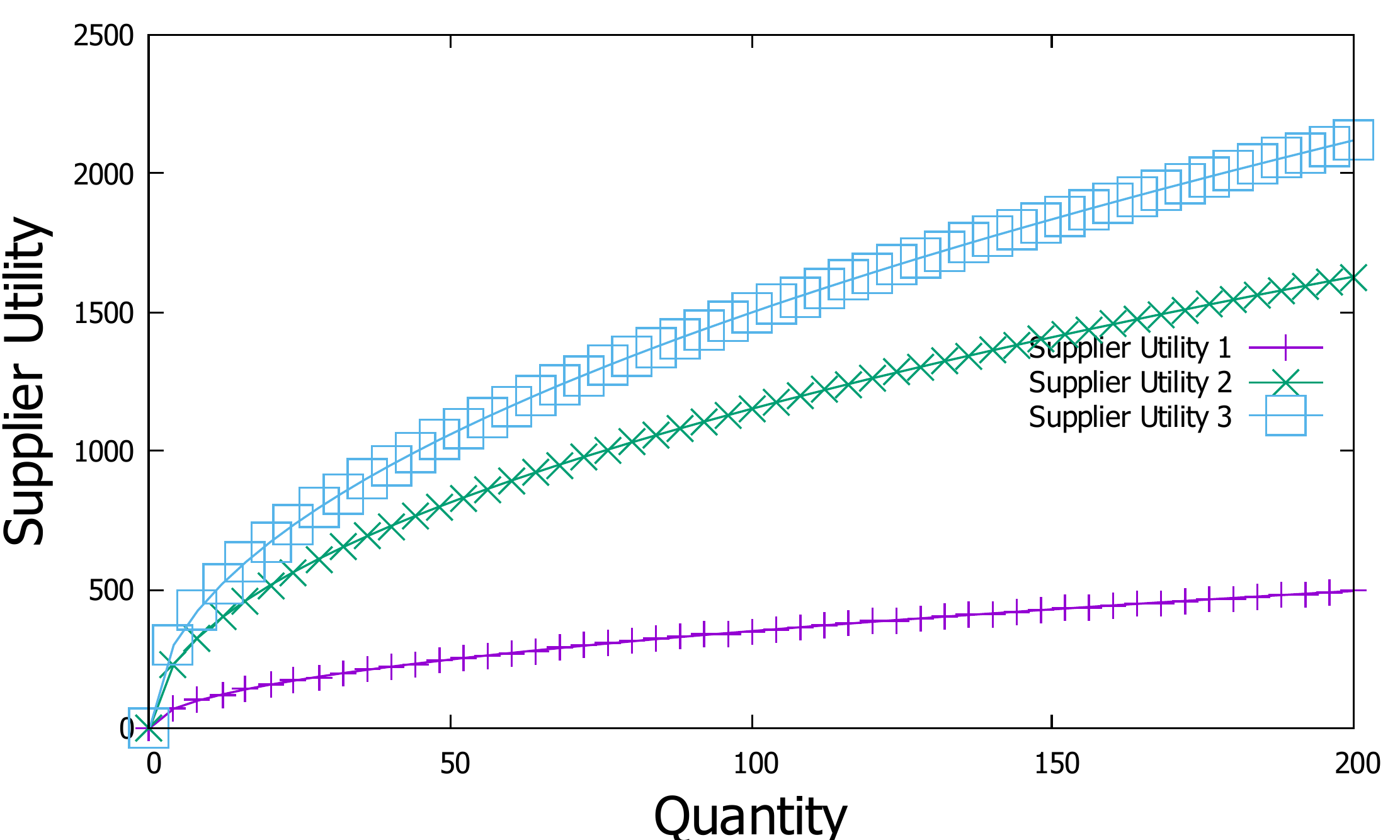}
\caption{Supplier utility function}
\label{fig:supplierutilitymonotonic}
\end{subfigure}
\begin{subfigure}{.50\textwidth}
\centering
\includegraphics[scale=0.38]{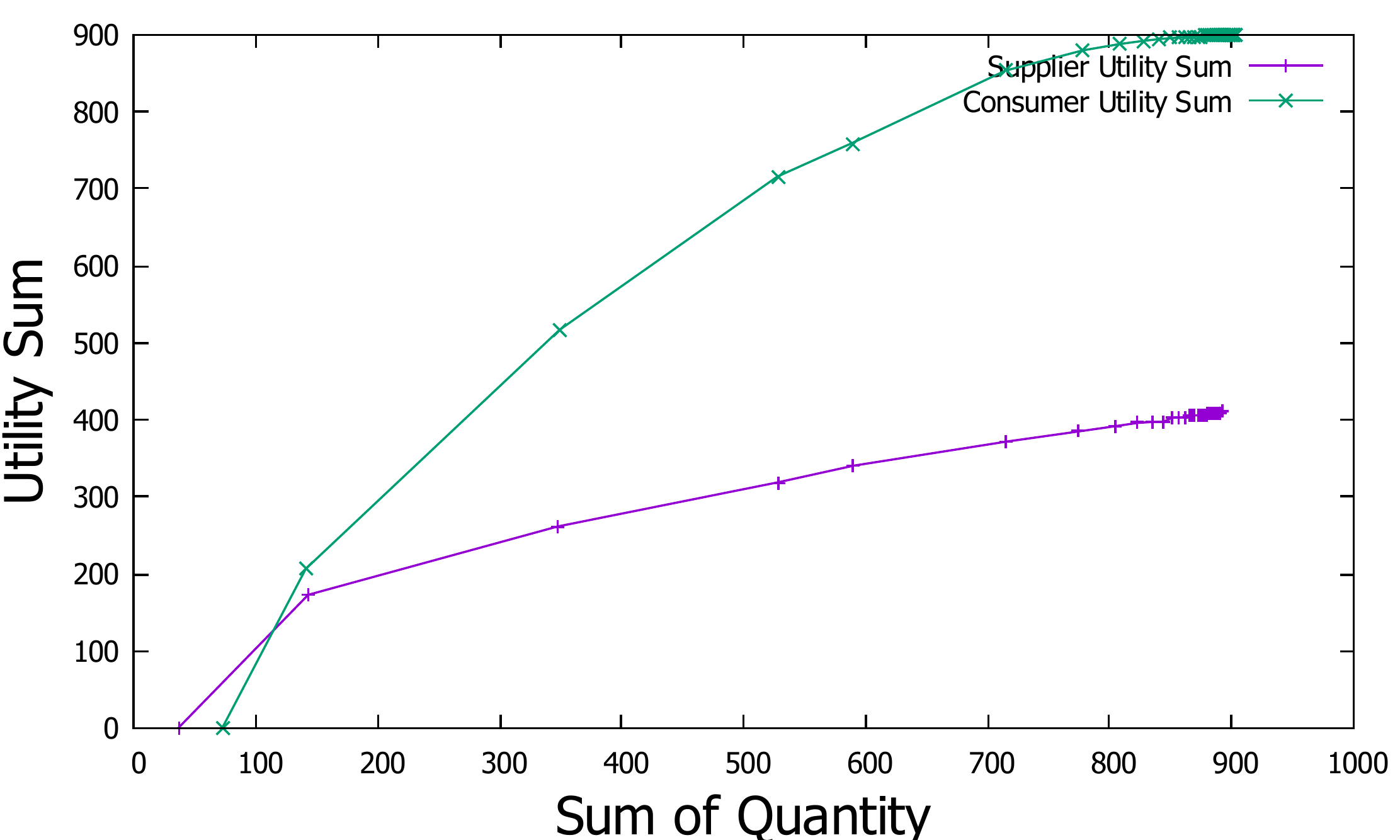}
\caption{Sum of Supplier and Consumer utilities with respect to the change in total supply/consumption averages}
\label{fig:utilitysumnonmonotonic}
\end{subfigure}
\begin{subfigure}{.50\textwidth}
\centering
\includegraphics[scale=0.38]{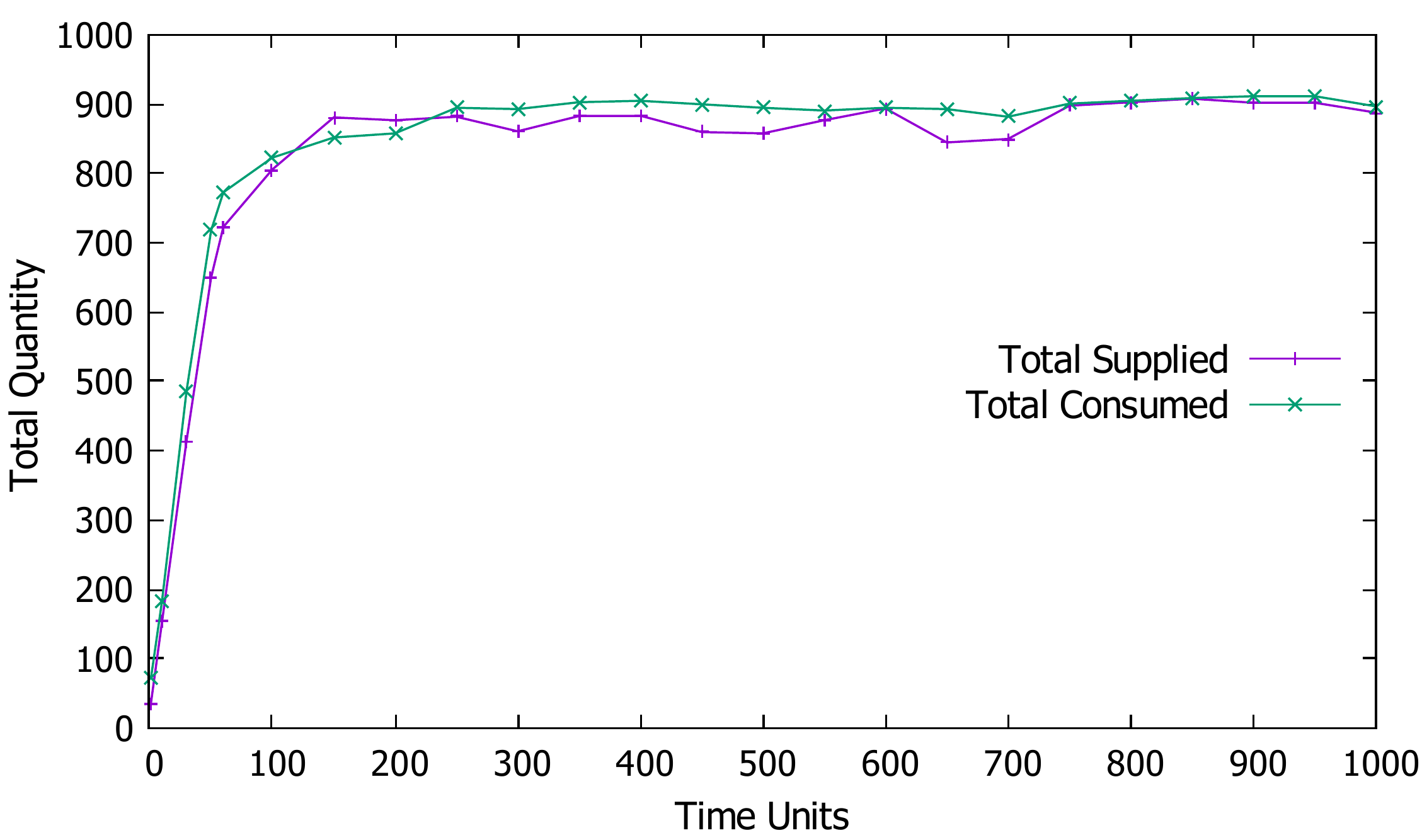}
\caption{Quantity supplied and consumed}
\label{fig:quantitynonmonotonic}
\end{subfigure}
\caption{Simulation Analysis for a monotonic supplier function, $f(x)={\ell_i}\,\sqrt[]{x}$}
\end{figure}

Now we consider a scenario where suppliers have a monotonic
non-decreasing utility function, while the consumers have concave
utility. So, the supplier utility (see
Figure~\ref{fig:supplierutilitymonotonic}) is of the form
\begin{equation}
{f_i}(\bar{x_i}(t)) = {\ell_i}\sqrt[]{\bar{x_i}(t)}
\end{equation}
while consumer utilities (Figure \ref{fig:18ConsumersUtility}) are of the form
\begin{equation}
{g_j}(\bar{y_j}(t)) = -\frac{(\bar{y_j}(t) - {y_j}^*)^2}{\hbar_j} + (1.5)\hbar_j
\end{equation}

On simulating for the same number of supplier and consumer agents,
along with the same constants as in the previous simulation, we
realize that sum of quantities supplied and consumed tend to saturate
around the optimal sum (see Figure~\ref{fig:quantitynonmonotonic}),
even when no particular maximum is present for the supplier utility. The
utility sum for consumers does converge towards the optimal sum (see
Figure~\ref{fig:utilitysumnonmonotonic}), unlike the supplier
utility's sum, as there is no particular maximum for the monotonic
non-decreasing functions.

\section{Conclusion}\label{sec-conclude}

In this paper, we have utilized the convergence properties of the AIMD
algorithm~\cite{aimd}, to solve the problem of
maintaining an equilibrium of supply and demand, for a group of
distributed agents \cite{enumula2009}, by also maximizing their
respective profits.

As showed in our simulations (see Section~\ref{simulation1}), for a
concave type utility functions for both supplier and consumer agents,
i.e., with utilities which have clear maxima, profit-maximization and
equal sum of supply and demand holds true.  By other simulations (see
Section~\ref{simulation2}), even when either side---supply or
demand---does not have a concave utility, equilibrium of supply and
demand is satisfied, along with the profit-maximization of the agent
with concave utility.

Considering the slew of applications requiring a global, dynamic
balance between supply and demand, such as the management of data
centers, energy producers and consumers connected to a smart grid, and
the like, we surmise that the work presented here can be put to
profitable use in several domains of application.

\section*{Acknowledgment}

Jia Yuan Yu was supported by the Natural Sciences and Engineering Research Council of Canada (RGPIN-2018-05096).


\bibliography{IEEEabrv,main}
\bibliographystyle{ieeetr}

\balance

\end{document}